\documentclass[twocolumn]{aastex6}

\usepackage{amsxtra}
\usepackage{amsmath}
\usepackage{amssymb}
\usepackage{graphicx}
\usepackage{txfonts}
\usepackage{color}
\usepackage{natbib}

\newcommand{\T}{\ensuremath{\Delta t}}

\newcommand{\blos}{\ensuremath{B_{\mathrm{LOS}}~}}
\newcommand{\mc}{MC{}}

\newcommand{\W}{\ensuremath{\textbf{W}}}
\renewcommand{\S}{\ensuremath{\mathbb{S}}}

\newcommand{\ov}{\ensuremath{b^{xy}}}
\newcommand{\wv}{\ensuremath{b}}

\newcommand{\sref}[1]{Section~\ref{#1}}
\newcommand{\eref}[1]{Equation~(\ref{#1})}
\newcommand{\fref}[1]{Figure~\ref{#1}}

\newcommand{\sunrise}{\textsc{Sunrise}}
\newcommand{\hmi}{\textsc{HMI\,}}

\newcommand{\imax}{\textsc{Sunrise II/IMaX{ }{}}}

\newcommand{\carcsec}{$\mbox{.\hspace{-0.5ex}}^{\prime\prime}$}

\DeclareMathOperator{\Tr}{Tr}

\begin{document}

\title{Maximum Entropy Limit of Small-scale Magnetic Field Fluctuations in the Quiet Sun}
\email{gorobets@leibniz-kis.de}

\author{
A.~Y.~Gorobets,$^{1}$
S.~V.~Berdyugina,$^{1}$
T.~L.~Riethm\"uller,$^{2}$
J.~Blanco~Rodr\'{\i}guez,$^{5}$
S.~K.~Solanki,$^{2,3}$
P.~Barthol,$^{2}$
A.~Gandorfer,$^{2}$
L.~Gizon,$^{2,4}$
J.~Hirzberger,$^{2}$
M.~van~Noort,$^{2}$
J.~C.~Del~Toro~Iniesta,$^{6}$
D.~Orozco~Su\'arez,$^{6}$
W.~Schmidt$^{1}$
V.~Mart\'{\i}nez Pillet,$^{7}$
\& M.~Kn\"olker,$^{8}$
}

\affil{
$^{1}$Kiepenheuer-Institut f\"ur Sonnenphysik, Sch\"oneckstr. 6, D-79104 Freiburg, Germany\\
$^{2}$Max-Planck-Institut f\"ur Sonnensystemforschung, Justus-von-Liebig-Weg 3, D-37077 G\"ottingen, Germany\\
$^{3}$School of Space Research, Kyung Hee University, Yongin, Gyeonggi, 446-701, Republic of Korea\\
$^{4}$Institut f\"ur Astrophysik, Georg-August-Universit\"at G\"ottingen, Friedrich-Hund-Platz 1, D-37077 G\"ottingen, Germany\\
$^{5}$Grupo de Astronom\'{\i}a y Ciencias del Espacio, Universidad de Valencia, E-46980 Paterna, Valencia, Spain\\
$^{6}$Instituto de Astrof\'{\i}sica de Andaluc\'{\i}a (CSIC), Apartado de Correos 3004, 18080 Granada, Spain\\
$^{7}$National Solar Observatory, 3665 Discovery Drive, Boulder, CO 80303, USA\\
$^{8}$High Altitude Observatory, National Center for Atmospheric Research, P.O. Box 3000, Boulder, CO 80307-3000, USA\\
}

\begin{abstract}

The observed magnetic field on the solar surface is characterized by a very complex spatial
and temporal behavior. Although feature-tracking algorithms have allowed us to deepen our
understanding of this behavior, subjectivity plays an important role in the identification and
tracking of such features.
In this paper, we continue studies \citep{paperI} of the temporal stochasticity of the magnetic field on the solar surface \textit{without} relying either on the concept of magnetic features or on subjective
assumptions about their identification and interaction.
We propose a data analysis method to quantify fluctuations of the line-of-sight magnetic field by means of reducing the temporal field's evolution to the \emph{regular} Markov process. We build  a representative model of fluctuations converging to the unique stationary (equilibrium) distribution in the long time limit with maximum entropy.
We obtained different rates of convergence to the equilibrium at fixed noise cutoff for two sets of data. This indicates a strong influence of the data spatial resolution and mixing-polarity fluctuations on the relaxation process.
The analysis is applied to observations of magnetic fields of the relatively quiet areas around an active region carried out during the second flight of the \textsc{Sunrise/IMaX} and quiet Sun areas at the disk center from the Helioseismic and Magnetic Imager on board the \emph{Solar Dynamics Observatory} satellite.
\end{abstract}


\maketitle

\section{Introduction} \label{section:intro}

The dynamics of magnetic fields in response to the surface turbulence seen as magnetic features in the solar photosphere is a challenging subject in the solar physics. Physically, features behave unlike rigid objects with sharp boundaries, so mutual interaction between them and between turbulent flows is ill-defined in the observations, which is worsened by the instrumental restrictions and artifacts.

Critical analysis of feature-tracking methods \citep{DeForestIV} motivated us to study physical properties
of the magnetic flux concentrations in the quiet Sun (QS), without relying on subjective assumptions about interaction and identification of the features.  We propose an alternative approach, which employs methods of statistical mechanics and thermodynamics for statistical characterization of temporal dynamics of the observable magnetic flux distribution, regardless of the flow configuration, particular events, or any other mechanisms that caused the observed distribution as defined in the magnetochemistry by \citet[][]{MagnetoChemistry}, see also \cite{DeForestI}. We apply the concepts of a statistical ensemble, its realizations and the stochasticity of (micro/macro-) states, in contrast to the feature-tracking and its calculus \citep[e.g.][]{DeForestI}.

Here, we demonstrate that the statistical complexity of the line-of-sight component (\blos) in the restless turbulent solar photosphere can relax to a canonical ensemble under certain idealizations, which we discuss.

By abandoning the concept of magnetic features and their ill-defined interactions \citep{DeForestIV}, we consider the magnetic field as a "flow" passing through a grid of detectors (image pixels). The grid of pixels provides sampling of this flow with a finite tempo-spatial resolution. Then, we define a \emph{state} of local flow to be the current value of \blos at a given time and space element (pixel). Every image pixel registers a time series of fluctuating states, which is interrupted by the noise; the noise is discarded. This consideration stems from the hydrodynamical experiments and leads to characterization of the magnetic field variations 
as being independent of (sub-) emergence mechanisms and all effects on account of advection, at least at this stage of our research.

It was shown by \cite{paperI} that temporal, pixel-recorded fluctuations of the quiet photosphere magnetic field permit a description as a space/time discrete Markov random variables, i.e. Markov chains (MCs). {A random variable is considered Makrov, if its future outcome is given only by the current value, i.e. it is independent of the past realizations.} The Markov properties were verified for longitudinal and transverse components, as well as their sum of squares, i.e. magnetic pressure variables.

The analysis was applied to the QS observations with a resolution of 0\carcsec{}15-0\carcsec{}18 and $33$~s cadence from the \textsc{IMaX} instrument on the first \sunrise~ mission \citep{Solanki2010, Solanki2016,Barthol_Imax,Gandorfer2011, Berkefeld2011, MartinezPillet2011}.

The Markov property implies that a stochastic process is completely described by probabilities independent of their distant past values (microstates) in time. That is to say, the future random outcome of \blos is given only by the \emph{current} value, i.e. there is no "memory" in the process.

A special class of MCs, so-called \emph{regular} Markov chains (RMCs), evolve in time toward configurations with increased entropy. Namely, an initial distribution of states tends to attain a time-independent limit, that is, there is a relaxation (thermalization) of a system described by a RMCs toward the equilibrium configuration with maximum entropy.

In the equilibrium, transitions between states in MC become statistically independent, i.e. they become independent even of the present state, so conditional state-to-state transition probabilities degenerate into trivial state occurrence probabilities.

The rate at which a system reaches the equilibrium distribution characterizes its "distance" from the equilibrium and thus can characterize a system's complexity. This is because higher organized state configurations have less entropy, i.e. less uncertainty. Qualitatively, a less disordered system dissolves the memory on the initial configuration slower, and thus stands further away from the equilibrium configuration in which any transition is equiprobable. In the language of thermodynamics, the above process of thermalization can be described by the following thought experiment. Consider a large volume of gas steadily perturbed away from its equilibrium macrostate and characterized by some (in general, time-dependent) probability density of microstates. At time $t_0$, we confine a smaller subvolume of the gas by impenetrable walls such that this subvolume (subsystem) becomes isolated from any interaction with the rest of the original volume (system). Then, according to the second law of thermodynamics, the subsystem eventually relaxes to the time-independent thermodynamical equilibrium macrostate (thermalization process).

The thermalization time of the subsystem counting from $t_0$ would depend on the nature of the fluctuations in the initial non-equilibrium macrostate and the physical properties of the system itself.

By analogy with this thought experiment, we analyze the relaxation time of the \blos fluctuations and its dependence on the spatial resolution of the observations considered as the system's free parameter. We propose and show in this paper that the complexity and rate of fluctuations depends on the spatial scale on which they occur: on larger scales magnetic structures interact (evolve) slower than on smaller ones. So, magnetic concentrations at higher resolutions "forget" (dissolve) their past configurations faster with respect to those at lower resolutions, and thus relaxation should run faster.

Physically, this can be explained by the difference in characteristic length scales of the photospheric turbulence revealed by different linear scales of the resolution elements.  At low-resolution, magnetic fields are instrumentally averaged on larger scales and thus a smoother structure is observed, which incorporates spatially more distributed fluctuations than at higher resolution. Consequently, the low-resolution pixel physically has more extended and thus more complex structure, which evolves slower since more time is required for turbulent motions to operate on larger scales to alter magnetic topology. In our analysis,  we show how spatial resolution influences the thermalization time.

The thermalization is an unobservable process but is theoretical scheme being used to obtain a quantitative measure of the observable \blos. During the thermalization, the observed Markov pairwise stochastic dependence between future and current \blos samples \emph{vanishes} for all possible pairs of pixel-recorded \blos samples, gathered in the entire field of view for the time of the observational campaign. Physically, the thermalization causes the \blos ensemble to have only statistical independent realizations, as well as samples forming them.

And more specifically, we aim to use the time required for thermalization as a characteristic measure of the fluctuations' complexity in the observed data, in the empirical model in which pixel-to-pixel \blos fluctuations have the possibility of becoming statistically independent in the long time limit. This empirical model of fluctuations conceptually is analogous to an isolated thermodynamical system, which we infer by reducing MCs in the data to the RMCs. In other words, we verify the existence of the \emph{unique} equilibrium distribution for the QS line-of-sight magnetic field density Markov fluctuations  under restrictions imposed by the RMCs.  We focus on the evolution of "extracted" RMCs from the data and estimate the rates of thermalization corresponding to the observations at different resolutions with comparable cadences.

The RMCs are built by the state's transitions of some particular properties. The mathematical conditions for an MC to be a regular one are mathematically more strict than those for the conventional Markov property. Thus, the requirements for stochastic data to follow RMC are not revealed by the Markov property test in \cite{paperI}. Therefore, applicability of the necessary conditions for RMC requires a dedicated study, which is presented in this paper. We study regularization of the Markov fluctuations for the observational data obtained at spatial resolutions of 0\carcsec15-0\carcsec18, 0\carcsec22 and 1\arcsec .

We use high-resolution \textsc{IMaX} data of comparatively quiet regions observed during the second \sunrise~ flight (referred to as \imax) and low-resolution QS data taken by the Helioseismic and Magnetic Imager on board the \emph{Solar Dynamics Observatory} satellite (\textit{SDO}/HMI). Note that \textit{SDO}/HMI was not available at the time of the first \textsc{Sunrise} flight, so that this analysis had to wait for the availability of new \textsc{IMaX} data from Sunrise.

We investigate qualitative and quantitative differences between the stochastic dynamics of \blos observed at different spatial scales with variable pixel selection criteria\footnote{The observational data from both instruments are parameterized by the threshold cutoff of noisy pixels, which leads to a number of data sets we work with (see Table~\ref{table}).} by comparison of the so-called long time limit behavior of the RMCs extracted from every data set.

In Section~\ref{section:observations} we describe the instruments and the observational data preprocessing, in \sref{subsection:theory} we provide a short introduction to the regular Markov random variables, and in \sref{subsection:estimates} we compare data at different spatial scales from the point of view of the RMCs. The results are discussed in Section~\ref{section:discussion}. Conclusions are presented in Section~\ref{section:conclusions}.

\section{Observational Data and Inference of Physical Parameters}\label{section:observations}

\subsection{Sunrise II/IM{\footnotesize A}X}

We use a time series recorded from 23:39 to 23:55~UT on 2013 June 12  during the second science flight of the balloon-borne solar observatory \sunrise{}
when the telescope pointed to the trailing part of the active region AR11768 (heliocentric angle $\mu=0.93$). For details of the 1\,m aperture telescope
and the gondola, we refer to \citet{Barthol_Imax}. Updates to the instrumentation operated during the second flight, as well as an overview of the flight and the obtained data, are given by \citet{Solanki2016}.
The influence of the disturbing Earth's atmosphere was minimized because 99\,\% of the airmass was below the telescope at the average flight altitude
of 36\,km. Image stabilization was achieved by using a tip/tilt mirror placed in the light distribution unit \citep{Gandorfer2011} and controlled by a correlating wavefront sensor \citep{Berkefeld2011}. The data
considered in this study were recorded with the Imaging Magnetograph eXperiment \citep[\textsc{IMaX};][]{MartinezPillet2011}, which scanned the Zeeman sensitive
Fe\,{\sc i} 5250.2\,\AA{} line (Land\'e factor $g=3$) at eight wavelengths (offset by $-120, -80, -40, 0, +40, +80, +120,$ and $+227$~m\AA{} from the line core) with a spectral resolution of 85\,m\AA{} and a cadence of $\T=36.5$\,s. \textsc{IMaX} had a field of view of $51\arcsec{} \times 51\arcsec{}$ at a plate scale of 0\carcsec{}0545 per pixel.

After the usual dark-field and flat-field correction, the data were demodulated and fringes and residual cross-talk were removed. An inflight phase-diversity
measurement allowed the determination of the point spread function (PSF), which was used to reconstruct the data. In contrast to the \textsc{IMaX} data
used by most other studies in this special issue \citep[whose reduction is described in][]{MartinezPillet2011,Solanki2016}, the data used in this study were not interpolated with respect to time prior to inversion (which is usually done
to correct for the influence of the solar evolution during the 36.5\,s \textsc{IMaX} cycle time) because the time interpolation is a weighted average of two \textsc{IMaX} cycles
and introduces a coupling between the data of successive cycles. After a 25\% global stray-light correction, the data were inverted with the SPINOR inversion code \citep{Frutiger2000a,Frutiger2000b} using a simple but robust atmospheric model: at three optical depth nodes (at $\log\tau=-2.5, -0.9, 0$) for the
temperature, and a height-independent magnetic field vector, line of sight velocity, and micro-turbulence. Maps of \blos show a noise level $\sigma=14\,\mathrm{Mx~cm^{-2}}$, determined as the standard deviation in small quiet regions.

\begin{figure}\centering \includegraphics[width=8cm]{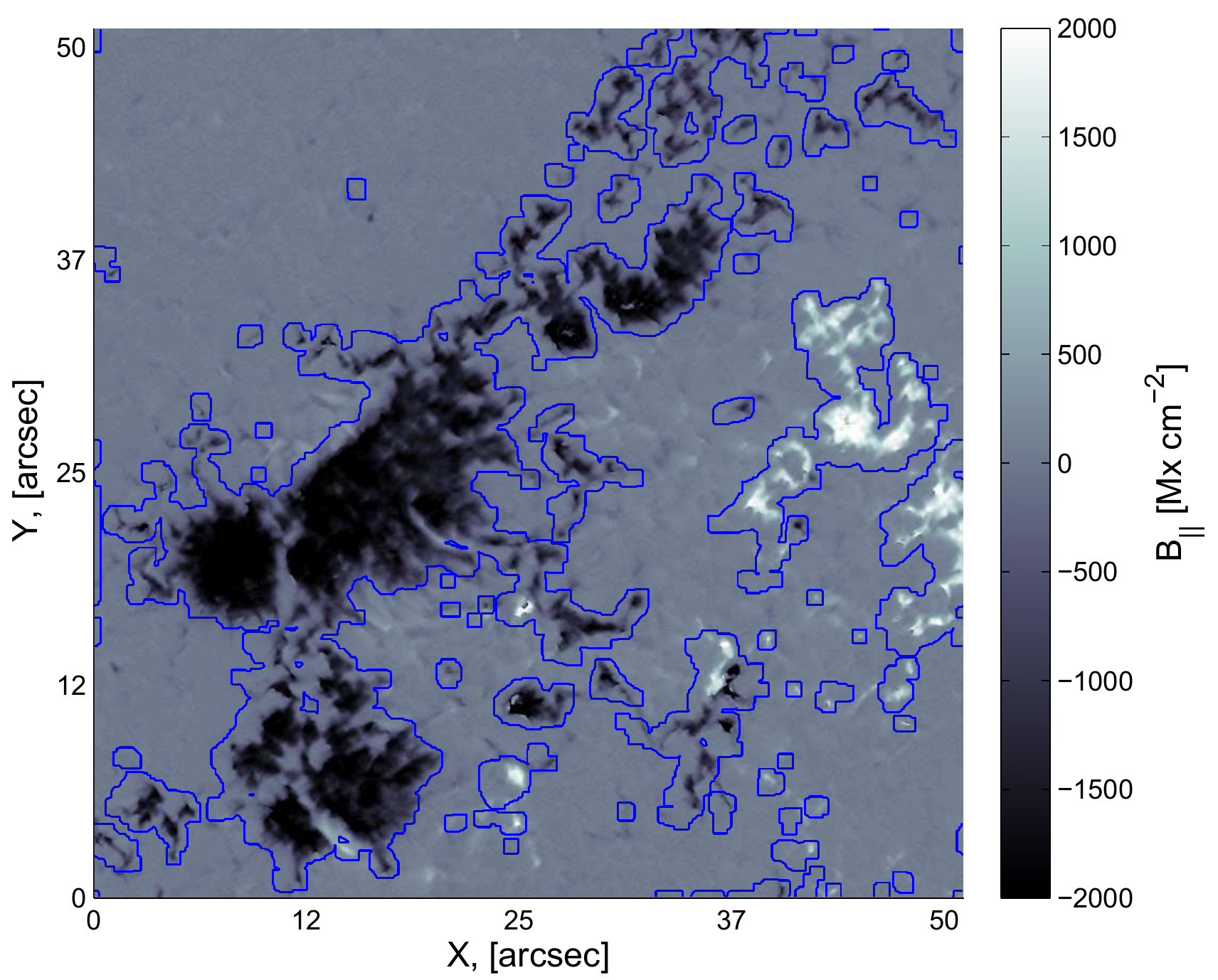} \caption{Snapshot of the \blos component of the magnetic field, from the \imax~ data set. The outlined regions designate the strong field areas and their boundaries, which are excluded from the analysis.}\label{figure:imax}\end{figure}

To be consistent with the analysis of \citet{paperI}, we limit the \imax~ data to have only regions of the relatively quiet photosphere as shown in \fref{figure:imax}. We exclude, from the whole $16\,\mathrm{min}$ series, pixels (and a 3-pixel surrounding) that present values higher than $\pm1.0\,\mathrm{k Mx~cm^{-2}}$ in even just one cycle of the observations.

{In the following, we refer to the \imax data as the one with median resolution of 0\carcsec165. We also analyze the same series but without PSF reconstruction, which has a resolution of 0\carcsec22 and $\sigma=7\,\mathrm{Mx~cm^{-2}}$.}

\subsection{\emph{SDO}/HMI}

To analyze magnetic field fluctuations at a low spatial scale, we consider uninterrupted HMI observations of the QS at the disk center, represented by a sequence of 9000 magnetograms in Fe {\footnotesize I} 6173 \AA{} line from 2015 November 28, 23:59:39~UT until 2015 December 4, 16:33:24~UT with a cadence of $\T=45~\mathrm{s}$ and resolution of $1\arcsec$ \citep{2012SoPh..275..207S}.

The images were preprocessed with the \texttt{hmi\_prep.pro} procedure from the \texttt{SolarSoft} package, and then cropped to the $500\times1000$ pixel area at the disk center.

To reduce "fast" oscillations in the \blos signal, the running local average is subtracted for three consecutive images, such that the local zero level is uniform in the data samples (see \eref{eq:train}).

To eliminate all possible induced ''memory effects'' by the data pipeline, we use the series \textit{hmi.M\_45s\_nrt}, which interpolates filtergrams linearly over three temporal intervals, instead of the $sinc$ spatio-temporal interpolation over five intervals in the regular (non-nrt) data series \citep{2011SoPh..269..269M, 2017ApJ...834...26K}.

The noise level of $\sigma=10.3~\mathrm{Mx~cm^{-2}}$ is used  according to estimates by \citet{SDO-tech} for the $45$\,s cadence magnetograms.

\section{Data analysis}\label{section:analysis}

In Section~\ref{subsection:theory}, we present a brief theoretical overview of RMCs.
In Section~\ref{subsection:estimates} we apply theoretical concepts to the statistical estimates inferred from the observational data.

\subsection{RMCs: theory and definitions} \label{subsection:theory}

The data analysis is based on the same assumptions as the technique by \citep{paperI}, which excludes tracking of the magnetic features.

Namely,  we consider pixels with a signal that is above the noise in three consecutive images simultaneously to form MC realizations (\eref{eq:train}). Technically, only evolving \blos above the noise cutoff for, at least, $2\T+3\times\text{exposure time}$ is analyzed, so the unknown structure of the weak fields below the noise cutoff and field disappearing into noise are excluded from our computations, since transitions from/to noise pixels to/from data pixels and noise to noise transitions are intentionally ignored.

Consequently, our view of \blos fluctuations has no field "emergence" from the void and "decay" to the noise level. Our attention now focuses solely on \blos variation at a pixel location rather than a specific event causing it. Thus, a specific state of the flow responsible for the advection is unnecessary to be specified.

We attribute apparent motions of the magnetic concentrations to general field fluctuations as well. In other words, if a transition $\wv_t \rightarrow \wv_{t+\T}$ at a given pixel is just due to trivial displacement of magnetic element, we still attribute this change in \blos  to a general field change (fluctuation), irrespective of the mechanism had caused it.

Following \cite{paperI}, consider a time and space-discrete stochastic Markov process (chain) $b(t)$
\begin{equation}
\ov_{t_1=t}\rightarrow\ov_{t_2=t+\T}\rightarrow\ov_{t_3=t+2\T}\cdots \;,
\label{eq:train}
\end{equation}
\noindent
where $\T$ is the cadence time (\sref{section:observations}), and $\ov$ is the observable
variable \blos inferred at image pixel $(x,y)$. The random variable $b$ is defined over a
finite set of all possible discrete values of $\ov$ (state space). The state space $\S$ has $S$ distinct elements representing equal bins of size $db$. The optimal value of $db$ is computed as in \citet{knuth} to be coarse-grained enough to smooth out a reasonable fraction of transitions due to noise. Nevertheless, the noise-induced transitions are impossible to filter out completely, since noise is always present in every data pixel as an additive random term, say $\xi$. Also, note that $\xi$ is zero mean Gaussian, and thus frequent transitions have zero contribution of $\xi$ on average.

Let $p=(p_i)_{i \in \S}$ be a probability density function (PDF) over the state space $\S$ such that $p_{i}db$ is the \textit{occurrence} probability of an observable $\ov$ in the state space bin $[i,i+db)$. The fixed bin size $db$ has been introduced for the estimation of the probabilities, and it is henceforth neglected in the equations for simplicity.

The $\T$-step conditional PDF $w_{ij}=w(b_{t+\T}=j|b_{t}=i)$ is defined such that
$w_{ij}$ is the probability for $b(t)$ to have value $j$ at time $t+\T$ if the random variable
$b(t)$ already had a value $i$ at earlier time $t$, with the normalization condition $\sum_{j \in \S} w_{ij}=1$.

The $S\times S$ transition matrix of one-step ($\T$) transition probabilities $w(\wv_{t+\T}|\wv_t)$,
\begin{equation}
\W :=[w_{ij}]_{i,j \in \S},
\end{equation}
\noindent
provides useful and important information on the temporal evolution of an MC. Namely, for a $2\T$ transition
\begin{align}
w(\wv_{t+2\T}&=j|\wv_{t}=i)  \\ \nonumber
&=\sum_k w(\wv_{t+2\T}=j|\wv_{t+\T}=k)w(\wv_{t+\T}=k|\wv_{t}=i) \\ \nonumber
&=\sum_k w_{jk} w_{ki} \\ \nonumber
&=\left(\W^2\right)_{ij\in\S}. \nonumber
\end{align}
\noindent
Then, by induction \citep[e.g.,][]{Privault}, for a general case we have
\begin{equation}\label{eq:power}\left[ w(\wv_{t+n\T}=j|\wv_t=i) \right]_{i,j \in \S} =\W^n, n\ge 0 .\end{equation}
\noindent
That is, a transition probability between states separated by an \textit{arbitrary} length in time is determined by one-time-step transition probabilities, which can be inferred from the data by considering two nearby images in the sequence.    

In practice, \eref{eq:power} offers remarkable diagnostic capabilities for the homogenous (time-independent transition probabilites) MCs using their transition matrices: its $n$th power allows computation of the probability of a jump of any length $n$ in time.

In general, transition probabilities change with the jump length $\W^m \neq \W^{m+1}$. However, for a large enough exponent $n$ the power of $\W$ stays practically the same as $n$ grows further.

In fact, in the long run ($n \rightarrow \infty$), RMC transition probabilities become independent of the initial state
\begin{equation}\label{eq:pi}\lim_{n \rightarrow \infty} w(b_{t+n\T}=j|b_{t}=i)=\pi_j ,\:\:{i,j \in \S}.
\end{equation}
Thus, the RMC is said to have a limiting PDF $\pi$, if the limit exists for all elements in $\S$, with the natural condition $\sum_{j\in\S} \pi_j=1$. The transition probability $w$ in the long time limit transforms into a simple probability of the final state. The limiting PDF $\pi$ is the \textit{unique} and \textit{stationary} (equilibrium) solution of the equation $\pi=\pi\W$, which leads to $\partial \pi/\partial t=0$.

This convergence imposes certain conditions on the temporal behavior of the MCs. Namely, an MC is regular if it is \textit{irreducible} and \textit{aperiodic}. We seek for these properties in the MCs obtained from the data.

The property of being irreducible means that every state of $\S$ is accessible from (communicates with) every other state of $\S$ , in a finite time (number of jumps). That is, phase space $\S$ cannot be split into (reduced to) disjointed subsets of states with no common elements.

The aperiodic chain revisits every state $i$ in $\S$ at random times $t$ such that the greatest common divisor (gcd) of these recurrence times for every state is $1$:
\begin{equation}\label{eq:gcd}
 t_{ii}=\left\{t\ge 1 :\: w(b_{t}=i|b_{0}=i) \right\},\:\,
\text{gcd} \left( t_{ii} \right)=1,\:\: {\forall i \in \S}.
\end{equation}\noindent
Otherwise, a chain with a periodic structure would have states with a recurrence times
being a multiple of some $t$.

The RMC with the limit \eqref{eq:pi} visits state $i$ on average once every $\langle t_{ii} \rangle$ steps: therefore,\begin{equation}\label{eq:prob-1} \pi_i= \langle t_{ii} \rangle^{-1}.\end{equation}

This is a fundamental relation between the limiting PDF and the state's average recurrence time \citep[for proof see e.g.] [p.83]{karlin}.

			
  	\subsection{Estimates} \label{subsection:estimates}
			

In the following, we  examine conditions for RMC and the existence of the stationary PDF for $16\,\mathrm{min}$ restored 0\carcsec165 and non-restored 0\carcsec22  \imax series and uninterrupted \emph{nrt} \hmi series of $\approx 4.7\:$-days long at a resolution of 1\arcsec.

The two \imax data cover the same solar region, so we can exclude the influence of the observed object morphology on our estimates.

Pixels strongly affected by noise are removed by the fixed $k\sigma$ threshold cutoff applied to the magnetograms before counting chain statistics. The integer $k$ is a free parameter in our study. It will be shown that the chain dynamics is strongly influenced by the pixel selection due to increasing $k$. Therefore, to assure the effect of a high cutoff level, we set $k$ up to an extreme value of, say $12\sigma$. In addition, the upper limit of $\pm1.5\,\mathrm{k Mx~cm^{-2}}$  for \blos is applied for \hmi data sets and $\pm1.0\,\mathrm{k Mx~cm^{-2}}$ for \imax.

It turns out that theoretical conditions for RMC, when applied to the data, work  like as a fine filtering of statistically insignificant states. The fraction of these states is less than 1\% for series with a rich statistical base, for which we assume that the regularity conditions are not underestimated (see Table~\ref{table}). We attribute the existence of such states to the different sorts of imperfections in the data, but first of all we attribute them to rare events inherent to statistical sampling, since these states appear predominantly in the tails of the PDF (see \fref{figure:gcd} and \ref{fig:DatMat}). Physically, the filtered-out rare events belong to the domain of strong fields, which is imposed by the nature of the QS, as seen by pixel-to-pixel analysis.

\subsubsection{Irreducibility}

\begin{figure}\centering\includegraphics[scale=1]{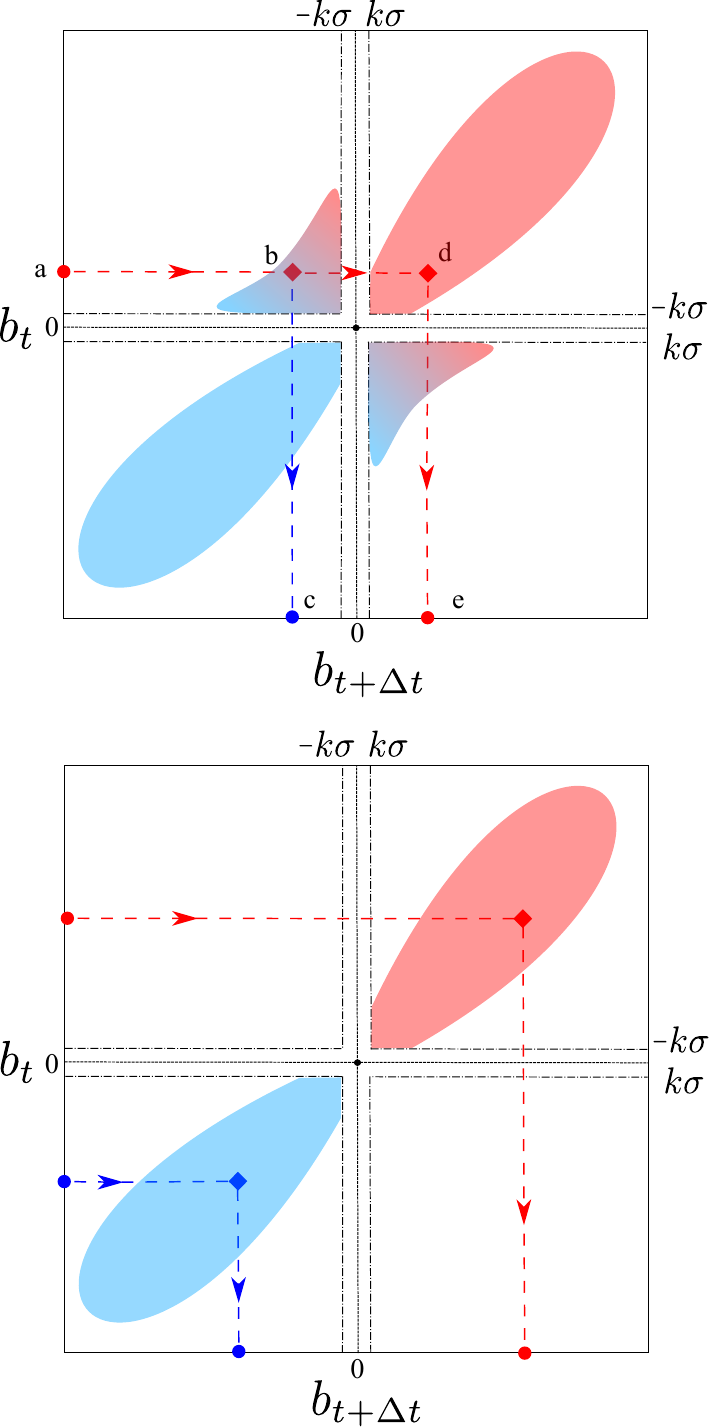}\caption{Schematic representation of the transition matrices obtained in the data analysis for irreducible MCs. {Top panel}: the scheme of the transition matrix for chains whose states communicate between states of different polarities. {Bottom panel}: the transition matrix of sign-preserving chains whose states do not communicate with states of opposite polarity. The dashed lines are examples of transitions. See the text for more details.}\label{fig:Wschem}\end{figure}

Qualitatively, the noise threshold cutoff governs the irreducibility of a chain: at low $k\sigma$, the chain transitions connect both polarities in \blos ($\pm\rightrightarrows\mp$ and $\pm\rightrightarrows\pm$), and at higher values of $k\sigma$, phase space splits into non-communicating subsets of opposite polarities (only $\pm\rightrightarrows\pm$). Therefore, corresponding transition matrices have different structures, which are schematically summarized in \fref{fig:Wschem}.

In \fref{fig:Wschem}, plume-like patches designate nonzero entries $w_{ij}$ of transition matrices with color-coding for polarities (red: positive, blue: negative). The dashed black lines show the $\pm k\sigma$ threshold cutoff around the zero level. In the top panel, patches with the gradient fill are the mixing (altering) polarity transitions.  For example, a sign-altering transition $\wv_t^{>0} \rightarrow \wv_{t+\T}^{<0}$ is shown by the path \textit{abc}, and its sign-preserving statistical counterpart for the same initial state $\wv_t^{>0}$ is shown by the path \textit{ade}. With increasing $k$, polarities stop communicating and transition matrices have a structure like the one sketched in the bottom panel of \fref{fig:Wschem}. In this case, all transitions are sign-preserving, as shown by the colored dashed lines.

Theoretically, irreducibility (all over communication in $\S$) has properties like the equivalence relation ($=$) i.e., (1) \textit{reflexivity} $i=i$, (2) \textit{symmetry} if $i=j$ then $j=i$, (3) \textit{transitivity} if $i=k$ and $k=j$ then $i=j$. Therefore, {to convert the transition matrix obtained from the data into the transition matrix representing the irreducible chains}, we remove all asymmetric elements with respect to the main diagonal of $\W$. In other words, if a transition $i\rightarrow i$ has no corresponding reverse transition $i\leftarrow i$ after collecting all possible transitions in the data set, we set such matrix elements to zero. This procedure guarantees satisfaction of all the properties of the mutual communication $\leftrightarrow$ (equivalence relation).

It is also possible to verify irreducibility explicitly, but at greater computational expense, by means of tracing  the overlapping subsets of the matrix elements, which had been attended by counting all available realizations of \eref{eq:train} in the data sets. This procedure had revealed the same results as those obtained by the method of removing the asymmetric transitions.

Physically, irreducibility is related to the ergodic properties of the fluctuations. Namely, if a particular observed value $\wv_t$ at time $t$ is never succeeded by $\wv'_{t+\T}$ at $t+\T$ (the transition $\wv_t\rightarrow\wv'_{t+\T}$ never occurs) it is always possible to construct a chain sample (not necessary observable in itself) such that $\wv$ and $\wv'$ will belong to the same realization in virtue of a intervening sequence of observed transitions  $\wv_t\rightarrow \wv''_{t+\T}\rightarrow\wv'''_{t+2\T}\rightarrow\cdots \wv'_{t+m\T}\rightarrow\cdots$. In other words, in irreducible MCs it is always possible to find a sequence of observable transitions between any two time-ordered states $\wv_{t_1}$ and $\wv'_{t_2}$, $t_1<t_2$ .

\subsubsection{Aperiodicity}

\begin{figure}\centering\includegraphics{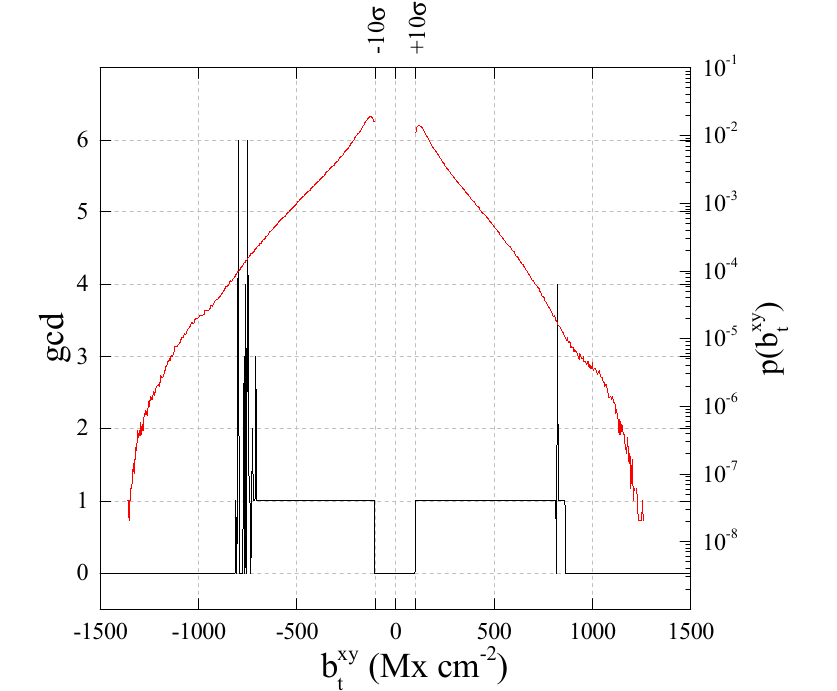}\caption{Aperiodicity of the states and occurrence probability density versus phase state variable for the case of an \hmi data set with a $10\sigma$ noise cutoff. $\ov_t$. {Black line}: the greatest common divisor estimated in the aperiodicity test. Red line: the state occurrence PDF $p=p(\ov_t)$.}\label{figure:gcd}\end{figure}

The aperiodicity is examined for states in every stochastic realization of \eref{eq:train} by counting occurrence times in units of $n\T$, where $n$ is the "local time" index. That is, every second sample $\ov_{t+n\T}$  in \eref{eq:train} is always counted at the local time $n=1$ in all realizations. 

The greatest common divisor is computed for every single state for all possible recurrence times from all observed realizations. The rows and columns of $\W$ with any number of aperiodic states ($\text{gcd}>1$) are set to zero. Those states that do not form transition pairs in \eref{eq:train} or that form insufficient short-length realizations, are considered to be of the undefined aperiodicity; for them we keep $\text{gcd}=0$, and they are discarded.

An example of an aperiodicity test is shown in \fref{figure:gcd}: the black line is the test for aperiodicity for HMI data with $10\sigma$ noise cutoff, and the red line is the state $\ov_t$  PDF $p=(p_i)=(p(\ov_t=i))$. Initially, gcd is set to zero, which is the blank value for all states. The test results in three types of gcd values: $\text{gcd}=0$ for undefined aperiodicity, $\text{gcd}>1$ for periodic states, and $\text{gcd}=1$ for aperiodic. The first two cases correspond to the excluded states from the analysis.

\begin{figure*}[ht]\centering \includegraphics[scale=1]{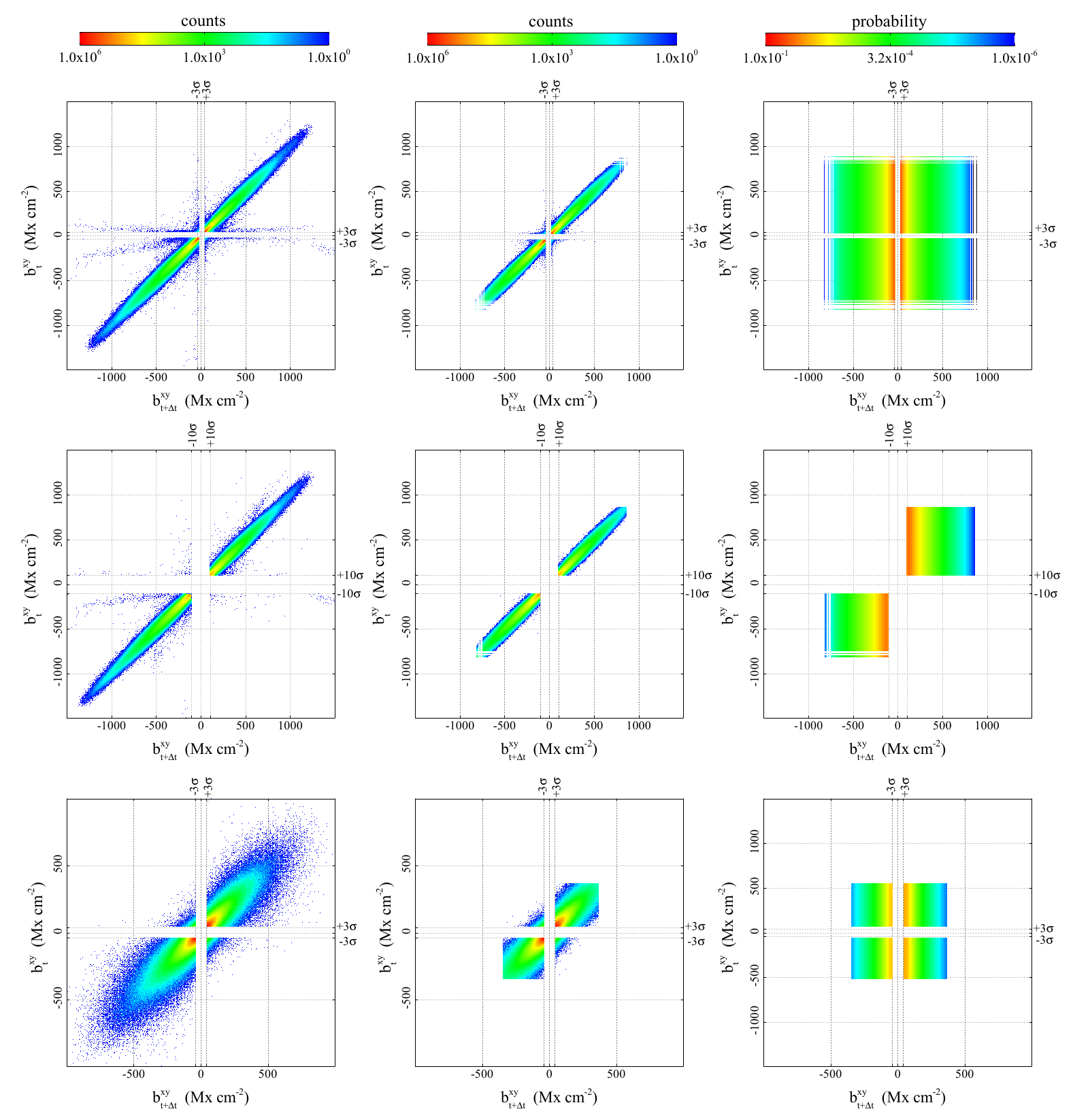} \caption{Transition and limiting matrices for \hmi data sets with $3\sigma$ (upper row), $10\sigma$ (middle row), and 0\carcsec165 \imax $3\sigma$ (bottom row) threshold cutoffs. { Left panels}: non-normalized transition counts from the data series. {Middle panels}: regularized transition matrices i.e. transitions between asymmetric and periodic states are removed. It is seen that sporadic and rare transitions in the tails are gone. {Right panels}: limiting matrices in which all rows are the same and columns are constants according to \eref{eq:matrix}. In the case of $10\sigma$, non-communicating submatrices should be considered as independent limits of the corresponding irreducible matrices, shown in the middle column, for each polarity separately. For the sake of simplicity, they are shown in the same grid.}\label{fig:DatMat}\end{figure*}

In the vicinity of zero, $\text{gcd}=0$, since there are no chain realizations in that region due to noise cutoff. In the tails $\text{gcd}=0$ with $p_i\neq 0$ and $w_{ij}\neq0$, due to undefined aperiodicity of the short-length realizations. In the test for aperiodicity, we search for repetitions of any state in a chain realization, and a state's aperiodicity becomes undefined when realizations are too short to contain any repetitions. A closer look at the changes of $\W$ due to regularization (compare the left and middle panels in \fref{fig:DatMat}) shows that we cut rare transitions along the borders of the cloud-shaped nonzero entries of the transition matrices. Apparently, such insignificant transitions present rare and short-length realizations, which cannot provide information about a state's aperiodicity.

The oscillating regions of the $\text{gcd}$ plot show periodic states whose occurrence probabilities are relatively low. These oscillating parts contain values of $\text{gcd}=1$ as well. In the case of the \imax~ data sets the $\text{gcd}$ estimation is \replaced{6: unreliable}{unreliable in the tails} due to shorter series of $16\,\mathrm{min}$. {Therefore, for all \imax~ data sets we set the gcd oscillating regions to zero until the first unity value in the continuous plateaus. This makes the data surely clean from possible underestimated long periods.}

\subsubsection{Convergence to the stationary limit}

The equilibrium PDF $\pi$ of an RMC can be computed from the matrix $\W^{n}$ in the long time limit. According to the theory \citep[e.g.][page 75]{Suhov}, $\W^{n\rightarrow\infty}$ should have all rows equal to each other, and these rows as well as the main diagonal are the PDF $\pi$:
\begin{equation}\label{eq:matrix}\lim_{n\rightarrow\infty}\W^{n}=\left[\begin{array}{cccc}\\[-2em]	\pi_1& \pi_2& \dots & \pi_S \\[0em]    \pi_1& \pi_2& \dots & \pi_S \\[0em]	\vdots& \vdots& \ddots & \vdots \\[0em]	\pi_1& \pi_2& \dots & \pi_S\end{array} \right].\end{equation}\noindent
Thus, it is sufficient to analyze convergence by checking the normalization condition for $\pi$ using the main diagonal elements of $\W^{n}$ (i.e. trace operator):
\begin{equation}\label{eq:limit}\lim_{n \rightarrow \infty}\Tr \W^n=\sum_j \pi_j =1\:.\end{equation}\noindent
So, when this limit is achieved at $n \rightarrow \infty$, the entire matrix converges to the unique PDF $\pi$ in the main diagonal and rows.

By raising transition matrices to a numerically large power $n$, we found for all data sets the existence of the theoretical limit \eqref{eq:limit}. In \fref{fig:DatMat}, we show examples of transition matrices at different stages in our analysis: the occurrence number of transitions with sporadic entries due to bad data (left column panels); irreducible matrices after application of the conditions for regularity (middle panels); and the corresponding limiting matrices (panels on the right). Note that due to plotting software the main diagonal of $\W$ shows a y-axis symmetric transformation in the schemes of \fref{fig:Wschem} and in \fref{fig:DatMat}.

In Table~\ref{table}, we demonstrate the effect of regularization of the observed chains. The table summarizes the percentage of transitions that had been removed by application of conditions for regularity (middle plots in \fref{fig:DatMat}). It turns out that \hmi data provide very reliable estimates, and less than $1\%$ transitions are discarded. We assume that with longer series of high-resolution data we would get similar results.

\begin{deluxetable}{c|ccc}\tablecaption{Percentage of removed transitions Due to Regularization of the MCs in the analyzed data sets.\label{table}}
\tablehead{\colhead{\small Cutoff Level}&\multicolumn{2}{c}{\imax} &\colhead{\hmi} \\ \colhead{$k\sigma$} & \colhead{0\carcsec165}&\colhead{0\carcsec22} & \colhead{1\arcsec}\\ \colhead{$k$}&\colhead{\%}&\colhead{\%} & \colhead{\%}}\startdata
$3$&  $3.00$& $15.64$  &$0.32$\\
$4$&  $4.51$ & $13.05$ &$0.46$\\
$5$&  $4.11$ & $15.43$ &$0.62$\\
$10$&-\tablenotemark{a}&-\tablenotemark{a}&$0.58$\\
$12$&-\tablenotemark{a}&-\tablenotemark{a}&$0.11$\enddata
\tablenotetext{a}{\footnotesize Not applied due to poor statistics.}
\end{deluxetable}

In both noise cutoff cases shown in \fref{fig:DatMat}, the limiting matrices $\W^{n\rightarrow\infty}$ (right panels) have rows that are equal to each other, which is in agreement with the theory (see \eref{eq:matrix}). These equal rows are the normalized equilibrium PDF $\pi$ (\eref{eq:pi} and (\ref{eq:limit})). This simple structure of constant values in columns of $\W^{n\rightarrow\infty}$ implies that every long time transition probability $w(\wv_{t_{n\rightarrow\infty}}=j|\wv_{t_0}=i)$ is independent of the initial state $\wv_{t_0}=i$. Remarkably, this is the key property of the equilibrium state: all transitions are equiprobable and have no memory of the initial configuration (states).

Hence, we found a subset of \blos fluctuations whose idealistic isolated evolution in the long time eventually reaches the stationary state, in which conditional probabilities $w(j|\cdot)$ will be replaced by simple occurrence probability $\pi(j)=\langle t_{jj}\rangle^{-1}$.

For the case of a sign-preserving \mc~ (right middle panel in \fref{fig:DatMat}), the limiting matrix has two independent submatrices that each have a structure as in \eref{eq:matrix} but for states of a single polarity. That is, the diagonal (and every row) of each submatrix is independently self-normalized as $n$ grows according to \eref{eq:limit}, just as the whole matrix for the sign-altering chain. Hence, the total sum of $\W$ is equal to 2 in this case.

\begin{figure}[ht]\includegraphics[scale=1]{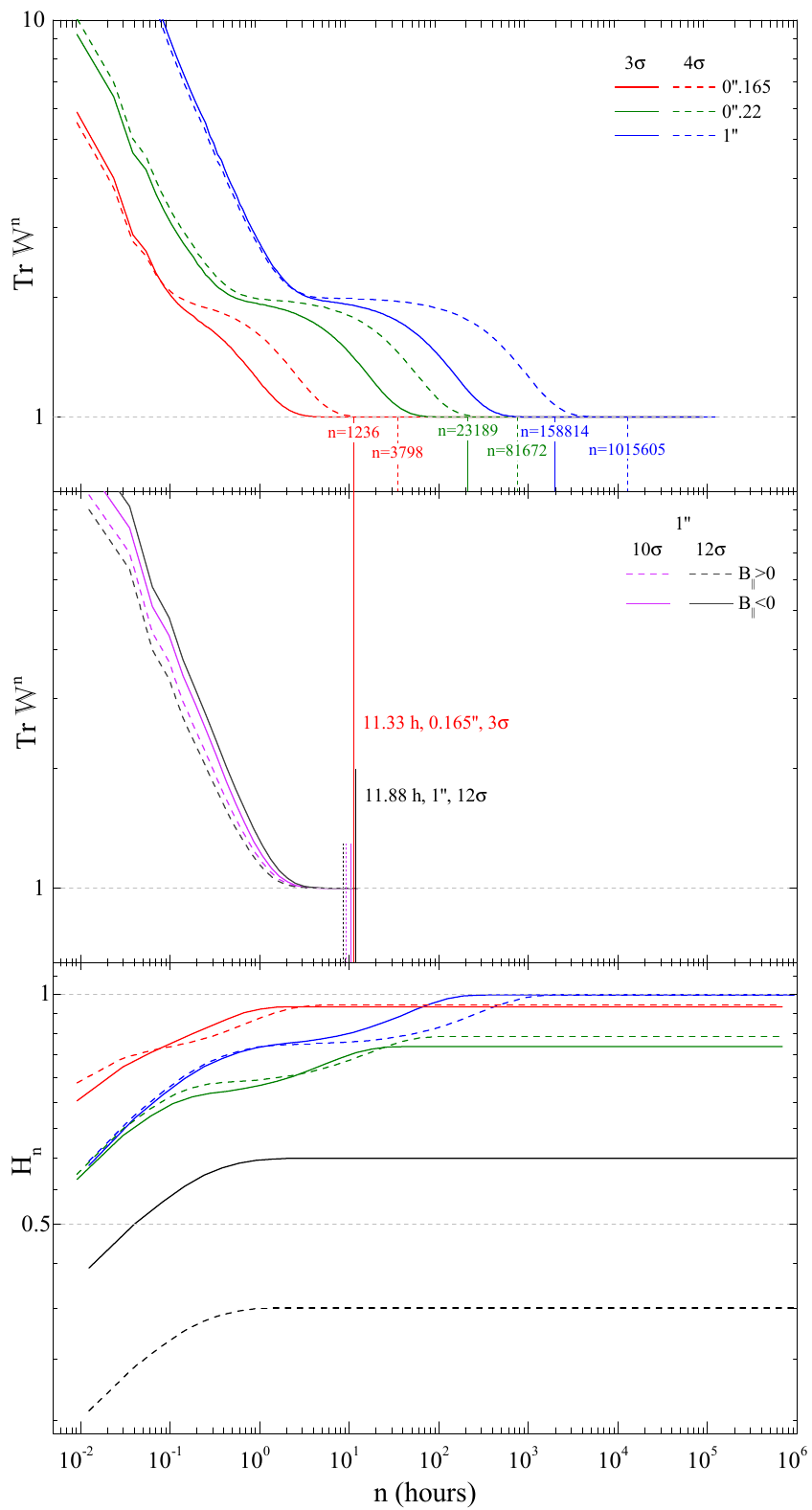}
\caption{Long time convergence. The trace operator in \eref{eq:limit} ({top} and {middle} panels) and conditional entropy $H_n$ in \eref{eq:entropy} ({bottom} panel) are plotted as a function of the matrix exponent $n$ (in time units). {Top}: cases of $3\sigma$ (solid lines) and $4\sigma$ (dashed lines) cutoff are shown for the data with sign-altering transitions. The color corresponds to the resolution: 0\carcsec165 (red), 0\carcsec22 (green), and 1\arcsec (blue).  The vertical lines indicate the minimal value of the exponents at which the prescribed precision of $10^{-7}$ is reached. Middle: the convergence of the transition matrices for the \hmi data with sign-preserving transitions due to high cutoff $10\sigma$ (purple) and $12\sigma$ (black). Transitions for negative \blos are shown by solid lines and positive ones are shown by dashed lines. To highlight fast convergence of RMC with sign-preserving polarities, the red line of high-resolution \imax data convergence is extended through the entire panel. Bottom: conditional entropy \eqref{eq:entropy} for the data sets plotted in the above panels, with corresponding color and line styles. The case of $10\sigma$ is abandoned to make the plot less crowded.}
\label{fig:convergence}\end{figure}

It turns out that the convergence rate toward the limiting PDF $\pi$ depends on the resolution and cutoff threshold. In \fref{fig:convergence}, we show the dependence of the main diagonal sum (trace operator) of $\W^n$ on power $n$, which is converted into time units of hours\footnote{With growing $n$ ( see \eref{eq:limit}) time changes on different scales for different instruments, since $\T$ for \hmi and \imax~ is	not the same.}. The top panel shows $\Tr(\W^{n\rightarrow\infty})$ for low threshold cutoffs of $3\sigma$ and $4\sigma$, corresponding to sign-altering RMC, and in the middle panel we plot the convergence of the transition matrices at high cutoffs of $10\sigma$ and $12\sigma$ for \hmi data when fluctuations of \blos do not change the polarity.

We say that function $f(n)=\Tr(\W^n)$ has converged to unity at equilibration (thermalization) time $t_{\mathrm{eq}}$, when
\begin{equation}
t_{\mathrm{eq}}=\min \{ n\ge 1:\, f(n)-1<10^{-7}\}.
\label{eq:precision}
\end{equation}
In other words, we seek precision, to at least six decimal places, for all cases shown in \fref{fig:convergence}.

In general, operator $\Tr(\W^{n})$ is a slowly decreasing function of $n$. For example, the exponent $n$ has to be about $10^6$ in the case of $4\sigma$ cutoff for \hmi data to achieve a particular equilibrium convergence limit of $\Tr(\W^{n})<10^{-7}$. In \fref{fig:convergence} {top}, for better illustration, every plot of $f(n)$ is accompanied by a corresponding vertical line of the same color and style to mark the value of $n$ at which the prescribed precision by \eref{eq:precision} is met for the first time.

Remarkably, single polarities at low-resolution converge as fast as mixed polarity high-resolution data at 0\carcsec{}165, with no strong dependence on the cutoff level.  In the middle panel of \fref{fig:convergence}, vertical lines mark the $10^{-7}$ convergence time corresponding to the independent polarities; they lie close to the red mark of the 0\carcsec{}165 \imax data with sign-altering transitions.

We summarize equilibration times in the Table~\ref{table:Times}, in which the larger $t_{\mathrm{eq}}$ are also presented in days. At lower resolutions with mixed polarity transitions, the equilibration time is longer because the structures are supposed to be more complex than those at higher resolution (see Section~\ref{section:discussion}).

\begin{deluxetable}{c|ccl}
\tablecaption{Equilibration times shown in Figure~\ref{fig:convergence}.\label{table:Times}}
\tablehead{\colhead{\small Cutoff Level}&\multicolumn{2}{c}{\imax} &\colhead{\hmi} \\
\colhead{$k\sigma$} & \colhead{0\carcsec165}&\colhead{0\carcsec22} & \colhead{1\arcsec}\\
\colhead{$k$}&\colhead{hr}&\colhead{hr} & \colhead{hr}}\startdata
$3$&  $11.33$& $212.57_{~8.9~(days)}$  &$1985.18_{~82.7~(days)} $\\
$4$&  $34.82$ & $748.66_{~31.2~(days)}$ &$12695.1_{~529~(days)}$\\
$10$&-\tablenotemark{a}&-\tablenotemark{a}&$9.36$\tablenotemark{\blos$>0$}\\
$~$ &                            &                            &$10.45$\tablenotemark{\blos$<0$}\\
$12$&-\tablenotemark{a}&-\tablenotemark{a}&$8.44$\tablenotemark{\blos$>0$}\\
$~$ &                            &                            &$11.88$\tablenotemark{\blos$<0$}
\enddata
 \tablenotetext{a}{\footnotesize Not applied due to poor statistics.}
 \end{deluxetable}

{To demonstrate the entropy growth to the maximum in the limit $n\rightarrow\infty$, we plot (see \fref{fig:convergence}, {bottom}) conditional entropy  \citep{6773024,Cover:2006:EIT:1146355}
\begin{align}\label{eq:entropy}
H_n(\ov_{t+n\T}=&j|\ov_{t}=i) =-\alpha^{-1}\sum_{i\in\S}p_i\sum_{j\in\S}w_{ij}\log(w_{ij})\:,
\end{align}\noindent
versus $n$, where $w_{ij}$ are given by the elements of the transition matrix $\W^n$. The normalization constant $\alpha=H_{n\rightarrow\infty}$ is the same conditional entropy but for the case of equilibrium state: $p_i$ is replaced by $\pi_i$ and $w_{ij}$ are taken from $\W^n$ with $n=10^9$. The non-decreasing functions $H_n$ emphasize a route of the thermalization processes toward the equilibrium state with maximum entropy. The convergence of $H_n$ to the unity in the limit of large $n$ characterizes how close the estimate of the occurrence probability $p$ is to the limiting PDF $\pi$, since due to normalization $H_n\rightarrow 1$ when $p_i\rightarrow\pi_i$. In the case of the $3\sigma$ \hmi data, this convergence is almost satisfied.

\section{Discussion}\label{section:discussion}

We studied whether it is possible to restrict Markov \blos fluctuations at different spatial scales in the quiet (and comparatively quiet  \imax~ case) Sun photosphere to a more restrictive class of the regular Markov processes. The answer for all data sets we have analyzed is positive: the vast majority of the observed transitions in fluctuating \blos is found to be consistent with the RMC.

The fundamental property of RMCs is their time evolution asymmetry: in the long run, they try to reach a unique and stationary (equilibrium) state distribution with maximum entropy.

It is found that a stationary limit exists for every combination of resolution and cutoff of the $11$ considered cases of the data parameterization (see Table~\ref{table}).

In a nutshell, the meaning of the "long run" term differs according to the spatial-scale-dependent structure of $\W$ and according to the noise level cutoff controlling the amount of mixing-polarity weak fields that enter the calculations. In the following, we try to analyze these factors one by one; however, they are hardly separated due to strong interdependence between them.

The rate at which the chain attains the stationary limit is much faster (up to 2 orders of magnitude) for fluctuations on smaller spatial scales with respect to the fluctuations on large scales with the same cutoff (see \fref{fig:convergence}, {top}). Note that $\sigma$ noise estimates in the data of $7, 10.3$, and $14\,\mathrm{Mx~cm^{-2}}$ are nearly comparable. The first crucial factor influencing $t_{\mathrm{eq}}$ is the structure of the transition matrix $\W$, specifically its spread along the diagonal that characterizes interrelations between states. For higher resolution, the spread in $\W$ is wider, indicating more random evolution of \blos.

The influence of a threshold cutoff on $t_{\mathrm{eq}}$ is, at least twofold, since it controls the amount of weak fields considered and the total presence of mixing-polarity transitions.

First, the relative amount of weak \blos fluctuating fields (with fast random dynamics) decreases with increasing noise cutoff level, and apparently the thermalization rate increases. Remarkably, this is the universal behavior for all resolutions  with mixing-polarity transitions.

Second, the strong \blos of single polarities of low resolutions converge to the limit contrastingly fast; the rate is of the same order (or even less) as the rate for small-scale mixing-polarity (see the middle panel in \fref{fig:convergence}). Thus, mixing-polarity transitions themselves play an essential role in modifying $t_{\mathrm{eq}}$.

Indeed, the influencing factor of the "weak fields amount" is \emph{inseparable} from that of the "presence of polarity-mixing transitions". This is because the observed polarity-altering transitions occur in the region of weak fields and almost all of the them are shaded by the noise level.

Therefore, it is difficult (with present resolutions and sensitivities) to infer whether sign alternation in the data is instrumental due to unresolved noisy transitions or if it is intrinsically of solar origin. Nevertheless, the noise removal keeps polarity-mixing transitions, albeit at a relatively low occurrence, and they appear as "bottleneck" transitions (with small but no vanishing probability, in the case of \hmi data) when in keeping states of different polarities in the same phase space. Hence, it should take more time for sign-altering fluctuations to overcome these constrains at equilibration, when all transitions should be equiprobable.

In contrast, at high-resolutions, the occurrence of polarity-mixing transitions increases and becomes more ingrained in the dynamics (notice the wider cloud around the diagonal in the bottom right panel of \fref{fig:DatMat}). Thus, we would describe them as the second main factor that decreases $t_{\mathrm{eq}}$ at high resolutions, when their presence becomes more statistically significant.

The observed mixing-polarity fields are used to argue for existence of the small-scale (fluctuating) dynamo  \citep[e.g.][]{Valentin}. If we assume that the observed pixel-to-pixel polarity-mixing transitions in our data are of solar origin, then this "imprint" of the dynamo action is the process speeding up the thermalization by a more rapid mixing of the states. Extrapolating the left column in \fref{fig:DatMat}, to higher resolutions and sensitivities (unavailable yet), the mixing-polarity transitions (and weak fields as well) could achieve a more distinct occurrence; giving more spherical symmetry to $\W$, i.e., blurring in greater amount its diagonal structure; thus, one can expect vanishing statistical difference between polarities and increasing overall mixing of the states ($t_{\mathrm{eq}}$ is {expected to be} very small in this case).

The observations of the QS \blos fluctuations with a sampling cadence of $33-45\,\mathrm{s}$ reveal a stochastic system at non-equilibrium. This follows from the transition matrices shown in the left and middle panels of \fref{fig:DatMat}, which have diagonally symmetric structure that is distinct from the equilibrium limit shown in the right panels of \fref{fig:DatMat} (see also \eref{eq:matrix})

\subparagraph{Biasing factors.} In general, spectro-polarimetric inversion techniques and various post-processing routines are able to induce memory artifacts on the temporal evolution of the observable \blos. To minimize such artifacts we use non-interpolated \imax data and \emph{nrt} \hmi series (see Section~\ref{section:observations}). In our method, the exact value of $t_{\mathrm{eq}}$ is sensitive to the noise estimates influencing the pixel selection, bin size $db$ and prescribed precision in \eref{eq:precision} as well. Thus, values in \fref{fig:convergence} should be read with a certain caution. The length of the data record influences statistics in the tails; for example, we expect to have more strong fields to be included in the calculation in case of longer \imax series for that domain.

\section{Conclusions.}\label{section:conclusions}

We propose a data analysis method to quantify fluctuations of the line-of-sight magnetic field. By means of reducing the temporal evolution of \blos to the regular Markov process, we build a representative fluctuating system  converging to the unique stationary (equilibrium) distribution and thus maximizing entropy in the long time limit. The method is applied to the QS observations at different spatial resolutions.

Different rates of maximizing entropy at fixed noise cutoff indicate that the high-resolution \blos fluctuations "forget" their initial distribution relatively fast compared to the low-resolution fluctuations. Thus, both qualitatively and quantitatively, we may  say that \imax~ observes fluctuations that are closer to an equilibrium state than the fluctuations we analyzed taken with the \hmi instrument.

Remarkably different rates of convergence to the long time limit are determined by the data-specific structure of the transition probabilities at different spatial resolution, in which the statistical significance of mixing-polarity transitions plays one of the main roles (if not the principal one).

While neglecting the effect of specific technical features on the instruments, different solar conditions and timing, we attribute the complexity of fluctuations, as well as its variability with the spatial resolution, to the nature of the overwhelming turbulence of the photospheric background.

We assume that the estimated relations between transition probabilities (i.e. the structure of the transition matrix) in our data sets arise from a universal property of the photospheric turbulence Thus, the relative proximity of the high-resolution \blos fluctuations to the equilibrium is also valid in the general context of the global properties of the turbulent photosphere.

\section*{Acknowledgements}
We acknowledge a partial support of this work by the European Research Council Advanced Grant HotMol (ERC-2011-AdG 291659).

The German contribution to \sunrise{} and its reflight was funded by the Max Planck Foundation, the Strategic Innovations Fund of the President of the Max Planck Society (MPG),
DLR, and private donations by supporting members of the Max Planck Society, which are all gratefully acknowledged. The Spanish contribution was funded by the Ministerio de Econom\'{\i}­a y Competitividad under Projects ESP2013-47349-C6
and ESP2014-56169-C6, partially using European FEDER funds. The HAO contribution was partly funded through NASA grant number NNX13AE95G. This work was partly supported by the BK21 plus program through the National Research Foundation (NRF)
funded by the Ministry of Education of Korea.

\textit{SDO} is a mission for NASA's Living With a Star (LWS) program. The \textit{SDO}/HMI data were provided by the Joint Science Operation Center (JSOC).


\end{document}